\documentclass[12pt]{article}
\usepackage{mathrsfs}
\usepackage{amsmath}
\usepackage{calrsfs}
\usepackage{mathrsfs}
\usepackage[mathscr]{euscript}
\usepackage{amsfonts}
\usepackage{amssymb, amsmath, cite}
\usepackage{color}

\newtheorem{theorem}{Theorem}
\newtheorem{lemma}{Lemma}
\newtheorem{proposition}{Proposition}

\newtheorem{corollary}{Corollary}

\newcommand\be{\begin{equation}}
\newcommand\ee{\end{equation}}
\newcommand\ber{\begin{eqnarray}}\newcommand\bea{\begin{eqnarray}}
\newcommand\eer{\end{eqnarray}}\newcommand\eea{\end{eqnarray}}
\newcommand\berr{\begin{eqnarray*}}
\newcommand\eerr{\end{eqnarray*}}

\newcommand{\lm}{\lambda}\newcommand{\bfR}{\mathbb{R}}\newcommand{\pa}{\partial}

\newcommand\re{\mathrm{e}}\newcommand\dd{\mbox{d}}

\newcommand{\ud}{\mathrm{d}}
\newcommand{\nm}{\nonumber}\newcommand{\nn}{\nonumber}

\newcommand{\ito}{\int_{\Omega}}

\newcommand{\vep}{\varepsilon}

\title{Resolution of Chern--Simons--Higgs Vortex Equations}

\author{Xiaosen Han\\Institute of Contemporary Mathematics\\School of Mathematics\\Henan University\\
Kaifeng, Henan 475004, PR China\\ \\
Chang-Shou Lin\\Taida Institute for Mathematical Sciences\\ National Taiwan University\\ Taipei, Taiwan 10617, ROC\\  \\Yisong Yang\\Department of Mathematics\\Polytechnic School of Engineering\\ New York University\\Brooklyn, New York 11201, USA
\\NYU--ECNU
Institute of Mathematical Sciences\\New York University - Shanghai\\3663 North Zhongshan Road, Shanghai 200062, PR China}
\date{}

\begin{document}
\maketitle

\begin{abstract}
It is well known that the presence of multiple constraints of non-Abelian relativisitic Chern--Simons--Higgs vortex equations makes it difficult to develop an existence  theory when the underlying Cartan matrix $K$ of the equations is that of a general simple
Lie algebra and the
strongest result in the literature  so far is when the Cartan subalgebra is of dimension 2. In this paper we overcome this difficulty by
implicitly resolving
the multiple constraints using a degree-theorem argument, utilizing a key positivity property of the inverse of the Cartan matrix deduced in
an earlier work of Lusztig and Tits, which enables a process that converts the
equality constraints to inequality constraints in the variational formalism. Thus this work establishes a general existence theorem
which settles a long-standing
open problem in the field regarding the general solvability of the equations. 
\medskip

\medskip 

\end{abstract}

 \section{Introdcution}
\setcounter{equation}{0}

The main result of this paper is a proof of
a general existence theorem for the doubly-periodic solutions of the
relativistic non-Abelian Chern--Simons--Higgs vortex equations whose Cartan matrix $K$ is that
of an arbitrary simple Lie algebra. The motivation of our work
originates from theoretical physics and development of new methods of mathematical analysis
to tackle systems of nonlinear equations of physical interests. Below
 we begin with a description of some of the field-theory aspects of our study.

{\em Field-theoretical origins.} In 1957, Abrikosov \cite{zAb} predicted in the context of the Ginzburg--Landau theory \cite{zdG,zGL,zSST,zT} that vortices of
a planar lattice structure
 may appear in
superconductors of the second type. Due to the complexity of the problem, a complete mathematical understanding of such
Abrikosov vortices has not been obtained yet, although in an over simplified situation, known as the
Bogomol'nyi--Prasad--Sommerfield (BPS) limit \cite{zBo,zPS}, an existence and uniqueness theorem is proved
under a necessary and sufficient condition  \cite{zWY} where the Abrikosov
lattice structure is realized through gauge-periodic boundary conditions conceptualized and formulated by 't Hooft \cite{ztH}
so that the governing elliptic equation is defined over a doubly periodic planar domain. Since early 1980, there has been a
growing interest in accommodating electrically charged vortices in condensed-matter physics \cite{zF,zFM1,FM,W2} such that
the presence of the Chern--Simons dynamics becomes imperative due to the Julia--Zee theorem which says that vortices in the classical Yang--Mills--Higgs models can only carry magnetism. For a recent review on these subjects and related literature, see \cite{zY}. It is relevant to describe such a
non-go theorem in the context of the simplest Yang--Mills--Higgs model here.

{\em The Julia--Zee theorem.} Consider the classical Yang--Mills--Higgs (YMH) model
in the adjoint representation of $SU(2)$ over
the Minkowski spacetime $\bfR^{2,1}$ of signature $(+--)$ and use $\mu,\nu=0,1,2$ to denote the spacetime coordinate indices.
Let ${\bf A}_\mu$ and $\phi$ be the gauge and Higgs fields written as isovectors, respectively.
Then the action density is
\be\label{x1.1}
{\cal L}=-\frac14 {\bf F}^{\mu\nu}\cdot {\bf F}_{\mu\nu}+\frac12 D^\mu\phi\cdot D_\mu\phi
-\frac\lm4(|\phi|^2-1)^2,
\ee
where the field strength tensor ${\bf F}_{\mu\nu}$ is defined by
$
{\bf F}_{\mu\nu}=\pa_\mu {\bf A}_\nu-\pa_\nu{\bf A}_\mu+e{\bf A}_\mu\times{\bf A}_\nu
$
and the gauge-covariant derivative $D_\mu$ is defined by
$
D_\mu\phi=\pa_\mu\phi+e{\bf A}_\mu\times\phi.
$
Here $\lm, e$ are positive coupling constants.
The Euler--Lagrange equations of (\ref{x1.1}) in the static limit are
\bea
D_i D_i \phi&=&-e^2(|{\bf A}_0|^2\phi-[\phi\cdot{\bf A}_0]{\bf A}_0)+\lm\phi(|\phi|^2-1),\label{x1.2}\\
D_j{\bf F}_{ij}&=&e({\bf A}_0\times {\bf F}_{i0}+\phi\times D_i\phi),\label{x1.3}\\
\Delta {\bf A}_0&=&-e(\pa_i[{\bf A}_i\times {\bf A}_0]+{\bf A}_i\times [\pa_i{\bf A}_0+e{\bf A}_i\times {\bf A}_0])
+e^2\phi\times({\bf A}_0\times\phi),\label{x1.4}
\eea
over the space $\bfR^2$, among which (\ref{x1.4}) is also the Gauss law constraint. The two-dimensional nature of these equations implies that their solutions may be interpreted as
the YMH vortices. On the other hand, the associated energy density, or Hamiltonian, of (\ref{x1.1}) may be calculated to be
\be\label{x1.5}
{\cal H}=\frac14 |{\bf F}_{ij}|^2+\frac12|\pa_i {\bf A}_0+e {\bf A}_i\times {\bf A}_0|^2+\frac12|D_i\phi|^2+\frac12e^2|{\bf A}_0\times\phi|^2+\frac\lm4(|\phi|^2-1)^2,
\ee
which is positive definite. In \cite{SY} it is proved that under the finite-energy condition
$
E=\int_{\bfR^2}{\cal H}\,\dd x<\infty,
$
any solution of (\ref{x1.2})--(\ref{x1.4}) must have a trivial temporal component for its gauge
field:
\be\label{x1.6}
{\bf A}_0={\bf 0}.
\ee
This result, first observed in \cite{JZ} for the Abelian Higgs model, later referred to as the
Julia--Zee theorem \cite{SY}, has an important physical implication.
Recall that, based on consideration on the manner of interactions,
't Hooft  \cite{'t} proposed that
the electromagnetic field $F_{\mu\nu}$ in the YMH theory is to be defined by the formula
\be\label{x1.7}
F_{\mu\nu}=\frac1{|\phi|}\phi\cdot {\bf F}_{\mu\nu}-\frac1{e|\phi|^3}\phi\cdot(D_\mu\phi\times
D_\nu\phi).
\ee
Hence, inserting (\ref{x1.6}) into (\ref{x1.7}), we obtain $F_{01}=F_{02}=0$, which indicates that there is no induced electric field in the model. In other words, the YHM vortices, Abelian
or non-Abelian, are purely magnetic
and electrically neutral.

{\em Emergence of the Chern--Simons type models.} The development of
theoretical physics requires the
presence of both electrically and magnetically charged vortices, also called
dyons, since such dually charged vortices have found applications in a wide range of areas including high-temperature superconductivity \cite{Kh,Ma}, optics \cite{Be},
the Bose--Einstein condensates \cite{In,Ka}, the quantum Hall effect \cite{So}, and superfluids
\cite{Ch,Ray,Sh}.
Thanks to the studies of Jackiw--Templeton \cite{JTemp}, Schonfeld \cite{Schonfeld}, Deser--Jackiw--Templeton \cite{DJT1,DJT2}, Paul--Khare \cite{PK}, de Vega--Schaposnik \cite{VS, VS1}, and Kumar--Khare \cite{KK}, starting from the early 1980's,
it has become accepted that, in order to accommodate electrically charged vortices, one needs to introduce into the action Lagrangian a Chern--Simons
topological term \cite{CS1,CS2}, which has also become a central structure in anyon physics \cite{FM,W1,W2}.
On the other hand, despite of the importance of electrically charged vortices with the added Chern--Simons dynamics,
it has been a difficult issue until rather recently \cite{zCGSY} to construct finite-energy solutions of the field equations because
of the indefiniteness of the action
functional as a consequence of the Minkowski signature of spacetime and the presence of electricity.
In 1990, it came as a fortune that Hong, Kim, and Pac \cite{zHKP} and Jackiw and Weinberg \cite{JW}
showed that when one uses only the Chern--Simons term and switches off the usual Maxwell term in the Abelian Higgs model one
can achieve a BPS structure and thus arrive at a dramatic simplification of the governing equations.
Subsequently, the ideas of \cite{zHKP,JW} were extended to non-Abelian gauge field theory models and a wealth of
highly interesting systems of nonlinear elliptic equations of rich structures governing non-Abelian Chern--Simons--Higgs vortices
was unearthed \cite{zD,zD2,zD3}. More recently, these ideas have also been further developed in supersymmetric gauge field theory
in the context of the Bagger--Lambert--Gustavsson (BLG) model \cite{zBL1,zBL2,zBL3,zBe,zCCR,zEMP,zG} and
the Aharony--Bergman--Jafferis--Maldacena (ABJM) model \cite{zabjm,CZZ,HYang,zkkkn} which have been the focus of numerous activities in
contemporary field-theoretical physics.

The present work is a complete resolution of the most general
relativistic Chern--Simons--Higgs vortex equations defined over a doubly periodic planar domain
with the Cartan matrix of an arbitrary simple Lie algebra. In the next section, we describe
the vortex equations and some of the key technical issues, including methodology. In the section
that follows, we state our
main existence theorem. In the subsequent sections, we prove this theorem.

\section{Vortex equations, technical issues, and methodology}
\setcounter{equation}{0}

Let $K=(K_{ij})$ be the Cartan matrix of a finite-dimensional semisimple Lie algebra $L$. We are interested in
the relativistic Chern--Simons--Higgs vortex equations  \cite{zD,zD2,zD3,Yang2,yang1} of the form
\be\label{a1}
\Delta u_i=\lm\left(\sum_{j=1}^n\sum_{k=1}^n K_{kj}K_{ji}\re^{u_j}\re^{u_k}-\sum_{j=1}^n K_{ji}\re^{u_j}\right)+4\pi\sum_{j=1}^{N_i}\delta_{p_{ij}}(x),\quad i=1,\dots,n,
\ee
where $n\geq1$ is the rank of $L$ which is the dimension of the Cartan subalgebra of $L$, $\delta_p$ denotes the Dirac measure concentrated at the point $p$, $\lm>0$ is a coupling constant, and
the equations are considered over a doubly periodic domain $\Omega$ resembling a lattice cell housing a distribution of point vortices located at $p_{ij},j=1,\dots,N_i, i=1,\dots,n$.
For an existence theory the ultimate goal  is to obtain conditions under which (\ref{a1}) allows or fails to allow a solution. In order to see the technical difficulties of the problem,
we take the beginning situation $n=1$ as an illustration for which the underlying gauge group may be either $U(1)$ or $SU(2)$ which is of fundamental importance in applications, such that (\ref{a1})
takes the scalar form
\be \label{zI2}
\Delta u=\lm\re^u(\re^u-1)+4\pi\sum_{j=1}^N\delta_{p_j}(x),\quad x\in\Omega.
\ee
Let $u_0$ be doubly periodic modulo $\Omega$ and satisfy $\Delta u_0=-\frac{4\pi N}{|\Omega|}+4\pi\sum_{j=1}^N\delta_{p_j}(x)$. Then $u=u_0+v$ recasts (\ref{zI2}) into
\be \label{zI3}
\Delta v=\lm\re^{u_0+v}(\re^{u_0+v}-1)+\frac{4\pi N}{|\Omega|},\quad x\in\Omega,
\ee
which is the Euler--Lagrange equation of the action functional
\be \label{zI4}
S(v)=\int_{\Omega}\left\{\frac12|\nabla v|^2+\frac\lm2\re^{2u_0+2v}-\lm\re^{u_0+v}+\frac{4\pi N}{|\Omega|} v\right\}\,\ud x.
\ee
It is easily seeing by taking constants as test functions that (\ref{zI4}) is not bounded from below over the space of doubly periodic functions. In order to tackle such a difficulty, it is  to take into
account of a natural constraint arising from integrating (\ref{zI3}). That is,
\be\label{zI5}
\int_{\Omega}\left(\re^{u_0+v}-\frac12\right)^2\,\ud x=\frac{|\Omega|}4-\frac{4\pi N}\lm,
\ee
which indicates that there fails to permit a solution when $\lm\leq\frac{16\pi N}{|\Omega|}$. At a first glance, it may be attempting to believe that a solution may be obtained by minimizing (\ref{zI4}) subject to
(\ref{zI5}). Unfortunately, there arises a Lagrange multiplier issue which prohibits a minimization process under the equality constraint (\ref{zI5}). To overcome this issue, we may take the decomposition
$v=c+w$ where $c\in\bfR$ and $\int_\Omega w\,\ud x=0$ and rewrite the constraint (\ref{zI5}) as
\be
\re^{2c}\int_\Omega \re^{2u_0+2w}\,\ud x-\re^c\int_\Omega\re^{u_0+w}\,\ud x+\frac{4\pi N}\lm=0,\label{xq}
\ee
which becomes a solvable quadratic equation in $\xi=\re^c$ if and only if the discriminant of (\ref{xq}) stays nonnegative,
\be \label{zI7}
\left(\int_\Omega \re^{u_0+w}\,\ud x\right)^2-\frac{16\pi N}\lm\int_\Omega \re^{2u_0+2w}\,\ud x\geq0,
\ee
so that $c$ may be represented as (say)
\be\label{zI8}
c=\ln\left(\int_\Omega \re^{u_0+w}\,\ud x+\sqrt{\left(\int_\Omega \re^{u_0+w}\,\ud x\right)^2-\frac{16\pi N}\lm\int_\Omega \re^{2u_0+2w}\,\ud x}\right)-\ln\left(2\int_\Omega \re^{2u_0+2w}\,\ud x\right).
\ee
Then it may be shown that a solution of (\ref{zI3}) can be obtained by minimizing the action functional
\be
I(w)=\int_{\Omega}\left\{\frac12|\nabla w|^2+\frac\lm2\re^{2u_0+2c+2w}-\lm\re^{u_0+c+w}\right\}\,\ud x+{4\pi N}c,
\ee
descending from (\ref{zI4}), subject to the inequality constraint (\ref{zI7}), with $c$ given by (\ref{zI8}), when $\lm$ is sufficiently large so that a minimizer occurs in the exterior of the constraint class which
rules out the Lagrange multiplier problem mentioned earlier. Finally, using the maximum principle and a continuity argument, it may be shown that there is a critical value of $\lm$, say $\lm_c>0$, such that there is no solution
when $\lm<\lm_c$ and solution exists when $\lm>\lm_c$. See \cite{CY1,yang1} for details. Later, it was shown in \cite{taran96} that there is solution at $\lm=\lm_c$ as well.
 More results on existence and asymptotic behavior of doubly periodic solutions  for \eqref{zI3} can be found in \cite{djlw,ntcv}.  Thus our understanding about the scalar case
(\ref{zI2}) or (\ref{zI3}) is fairly satisfactory.

When $G=SU(3)$ so that $n=2$, things already become rather complicated because now one needs to resolve two coupled quadratic constraints. This more complicated problem was studied by Nolasco and Tarantello \cite{nota} who refined and improved
the inequality-constrained minimization method developed in \cite{CY1} for the $n=1$ situation and showed  that solutions in this $n=2$ situation exist as well when $\lm$ is sufficiently large.
Note that, since in this case we are treating a system of equations, for which the maximum principle cannot be used and thus a continuity argument as that in \cite{CY1} is not available, an existence result under the condition that
$\lm$ is sufficient large may be the best one can hope for. See \cite{CHLS} for some new development regarding more generalized $2\times 2$ systems arising in
the Chern--Simons theory.

The  contribution of the present article is a successful settlement of the situation when the gauge group
$G$ is any compact group and in particular the Cartan matrix $K$ of the equations is that of
an arbitrary simple Lie algebra, of rank $n$. The system now consists of $n$ nonlinear equations and results in $n$ quadratic constraints,
which cannot be resolved explicitly as in \cite{CY1,nota}. In order to unveil the constraints difficulties we encounter and the varied levels of effeciency of the implicit constraint-resolution
methods we use, we shall present a special method that works for the $SU(4)$ situation, which has been extended to tackle the $SU(N)$ situation in \cite{haya}, and then
a general method that works for all situations. The special method may be described as a ``squeeze-to-the-middle" implicit-iteration strategy whose validity depends on
the structure of the Cartan matrix of $SU(N)$. The general method, on the other hand, uses a degree-theorem argument, which does not depend on the detailed specific
numeric structures of
the Cartan matrix.  Rather, we shall see that, for a simple Lie algebra (say), things may be worked out miraculously to ensure an acquirement of all  the needed {\em apriori} estimates
so that the multiple quadratic constraints allow an implicit resolution in any situation under consideration.

\section{Chern--Simons--Higgs equations and existence theorem}
\setcounter{equation}{0}

Our purpose is to  carry out a complete resolution of  the existence of doubly periodic solutions to \eqref{a1} with very general Cartan matrix $K$.
In order to treat the system in a unified   framework, we need some suitable assumption on the matrix $K$. For a semi-simple Lie algebra, we know that the associated Cartan matrix has the property: the  diagonal entries $K_{ii}$ assume the same  positive integer 2, all  off-diagonal entries $K_{ij}$ $(i\neq j)$ can only assume the  non-positive integers
$-3,-2,-1,0$, and  $K_{ji}=0 $ if $K_{ij}=0$.   This motivates us to consider (\ref{a1}) with a general matrix $K$, which  satisfies
\ber
 K^\tau&=&PS, \label{a2}
\eer
where  $P$ is a diagonal matrix with
 \ber
 P&\equiv&{\rm diag}\{P_1, \dots, P_n\},  \quad   P_i>0, \quad i=1, \dots, n,\label{a3}
 \eer
 $S$ is a  positive definite matrix of the form
 \ber
S\equiv\begin{pmatrix}
 \alpha_{11}&-\alpha_{12}& \cdots &\cdots&\cdots&-\alpha_{1n}\\
 \vdots& \vdots& \vdots &\vdots& \vdots&\vdots\\
 -\alpha_{i1}& -\alpha_{i2} &\cdots&\alpha_{ii}&\cdots&-\alpha_{in}\\
 \vdots& \vdots& \vdots &\vdots&\vdots& \vdots\\
 -\alpha_{n1}& -\alpha_{n2} &\cdots&\cdots&-\alpha_{nn-1}&\alpha_{nn}
\end{pmatrix},\label{a4}
\eer
\ber
&&\alpha_{ii}>0, \quad i=1, \dots, n,\quad \alpha_{ij}=\alpha_{ji}\ge0, \, i\neq j=1, \dots, n,\label{a5z1}\\
 && \, \text{and all the entries of} \quad S^{-1}\,  \text{ are positive}. \label{a5z4}
\eer

In fact the assumptions on the matrix on $K$ are broad enough to cover all simple Lie algebras realized  as
$A_n, B_n, C_n, D_n, E_6, E_7, E_8, F_4,G_2$ \cite{Hum,kac}, thanks to the work of Lusztig and Tits \cite{Luti}.

 As a result of \eqref{a5z4},   the entries of  $(K^\tau)^{-1}=S^{-1}P^{-1}$ are all positive. In particular, we have
 \ber
 R_i\equiv \sum\limits_{j=1}^n((K^\tau)^{-1})_{ij}>0,\quad  i=1, \dots, n.\label{a6}
 \eer

Here is our   main  existence theorem  for \eqref{a1}.

\begin{theorem}\label{th1}
Consider the non-Abelian Chern--Simons system \eqref{a1} over a doubly periodic domain $\Omega$ with the matrix $K$ satisfying \eqref{a2}--\eqref{a5z4}.   Let   $p_{i1}, \dots, p_{iN_i}$
$ ( i=1, \dots, n)$ be any given points on $\Omega$, which need not to be distinct.
Then there hold the following conclusions.

{\rm (i) } (Necessary condition) If the system \eqref{a1} admits a solution, then
 \ber
  \lambda> \lambda_0\equiv\frac{16\pi}{|\Omega|}\frac{\sum\limits_{i=1}^n\sum\limits_{j=1}^nP_i^{-1}(K^{-1})_{ji}N_j}{\sum\limits_{i=1}^n\sum\limits_{j=1}^nP_i^{-1}(K^{-1})_{ji}}. \label{a5a}
 \eer

{\rm (ii) } (Sufficient condition)  There exists a large  constant $\lambda_1>\lambda_0$ such that when $\lambda>\lambda_1$ the system \eqref{a1} admits a solution $(u_1^\lambda, \dots, u_n^\lambda)$.

{\rm (iii) } (Asymptotic behavior)  The solution  $(u_1^\lambda, \dots, u_n^\lambda)$ of \eqref{a1} obtained in { \rm (ii) }  satisfies
\ber
 \ito(\re^{u_i^\lambda}-R_i)^2\ud x\to 0\quad \text{as} \quad \lambda \to \infty, \quad i=1, \dots, n,\label{a5b}
\eer
 where $R_i$ $ ( i=1, \dots, n)$  are  defined by \eqref{a6}.

 {\rm (iv) } (Quantized integrals)  The solution  $(u_1^\lambda, \dots, u_n^\lambda)$ of \eqref{a1} obtained in {\rm(ii)} possesses  the following quantized integrals
  \be
    \ito \left(\sum_{j=1}^n K_{ji}\re^{u_j^\lambda}-\sum_{j=1}^n\sum_{k=1}^n K_{kj}K_{ji}\re^{u_j^\lambda}\re^{u_k^\lambda}\right)\ud x=\frac{4\pi N_i}{\lambda}, \quad i=1, \dots, n.\label{a5c}
  \ee

\end{theorem}

In the subsequent sections we prove the theorem.

\section{Necessary condition and variational formulation }
 \setcounter{equation}{0}
In this section we  shall  find  a necessary condition for the existence  of  solutions  of \eqref{a1} and present  a variational   formulation.

For convenience, we first use the translation
\be
u_i\to u_i+\ln R_i, \quad   i=1, \dots, n, \label{a7}
\ee
to recast the system  \eqref{a1} into a normalized form:
 \be
\Delta u_i=\lm\left(\sum_{j=1}^n\sum_{k=1}^n \tilde{K}_{jk}\tilde{K}_{ij}\re^{u_j}\re^{u_k}-\sum_{j=1}^n \tilde{K}_{ij}\re^{u_j}\right)+4\pi\sum_{j=1}^{N_i}\delta_{p_{ij}}(x),\quad i=1,\dots,n\label{a8}
\ee
whose  vector version reads
\ber
 \Delta \mathbf{u}=\lambda\tilde{K}\mathrm{U}\tilde{K}(\mathbf{U}-\mathbf{1}) +4\pi\mathbf{s}, \label{a9}
\eer
where  the notation
\ber
 && \tilde{K}\equiv K^\tau R=PSR, \quad R\equiv{\rm diag}\big\{R_1, \dots,  R_n\big\}, \label{a10}\\
  &&\mathbf{u}\equiv(u_1, \dots, u_n)^\tau, \,  \mathrm{U}\equiv{\rm diag}\big\{\re^{u_1},  \dots, \re^{u_n}\big\},\, \mathbf{U}\equiv(\re^{u_1},  \dots, \re^{u_n})^\tau, \label{a11}\\
&& \mathbf{1}\equiv(1, \dots, 1)^\tau,\quad \mathbf{s}\equiv\left(\sum\limits_{s=1}^{N_1}\delta_{p_{1s}},  \dots,   \sum\limits_{s=1}^{N_n}\delta_{p_{ns}}\right)^\tau,\label{a12}
\eer
will be observed throughout this work.
Note that, since the matrix $S$ is positive definite, so are  the matrices
 \ber
  A \equiv P^{-1}S^{-1}P^{-1} \quad \text{and}\quad Q\equiv RSR. \label{a6a}
 \eer

To solve  \eqref{a8} or \eqref{a9} over  a doubly periodic domain, we need to introduce some background functions to remove the Dirac  source terms.
Let $u_i^0$ be the solution of the following problem \cite{aubi}
\berr
 \Delta u_i^0=4\pi\sum\limits_{s=1}^{N_i}\delta_{p_{is}}-\frac{4\pi N_i}{|\Omega|},\quad \ito u_i^0\ud x=0,
\eerr
and $u_i=u_i^0+v_i, \, i=1,  \dots,  n$.   In the sequel we will use the $n$-vector  notation
\be
 \mathbf{v}\equiv(v_1, \dots, v_n)^\tau, \quad \mathbf{N}\equiv(N_1, \dots, N_n)^\tau, \quad \mathbf{0}\equiv(0, \dots, 0)^\tau. \label{a13}
 \ee
Thus  the  system  \eqref{a8} or  \eqref{a9} becomes
 \be
\Delta v_i=\lm\left(\sum_{j=1}^n\sum_{k=1}^n \tilde{K}_{jk}\tilde{K}_{ij}\re^{u_j^0+v_j}\re^{u_k^0+v_k}-\sum_{j=1}^n
\tilde{K}_{ij}\re^{u_j^0+v_j}\right)+\frac{4\pi N_i}{|\Omega|},\quad i=1,\dots,n,\label{a14}
\ee
or
\ber
 \Delta \mathbf{v}=\lambda\tilde{K}\mathrm{U}\tilde{K}(\mathbf{U}-\mathbf{1})+ \frac{4\pi \mathbf{N}}{|\Omega|},   \label{a15}
\eer
 where
\ber
\mathrm{U}={\rm diag}\big\{\re^{u_1^0+v_1},  \dots, \re^{u_n^0+v_n}\big\}, \quad \mathbf{U}=(\re^{u_1^0+v_1}, \dots, \re^{u_n^0+v_n})^\tau. \label{a16}                                                                                                                                                                                                     \eer

We now unveil a necessary condition for the existence of solutions of \eqref{a1}.  To this end, we  rewrite   \eqref{a15}, after multiplying both sides of \eqref{a15} by $A$,  equivalently as
\ber
 \Delta A \mathbf{v}&=&\lambda\mathrm{U}Q(\mathbf{U}-\mathbf{1})+ \frac{\mathbf{b}}{|\Omega|}, \label{a17}
\eer
where $A$ and $Q$ are defined  in  \eqref{a6a}, and
 \ber
  \mathbf{b}\equiv(b_1, \dots, b_n)^\tau\equiv 4\pi A\mathbf{N}= 4\pi P^{-1} S^{-1}P^{-1} \mathbf{N}. \label{a18}
 \eer
Noting \eqref{a5z4},  we obtain
 \be
  b_i>0, \quad i=1, \dots, n, \label{a19}
 \ee
which will be used in the sequel.

Let $\mathbf{v}$ be a solution of  \eqref{a15}, which is  of course  a solution of \eqref{a17}.     Then  integrating  \eqref{a17} over $\Omega$,  we obtain the natural constraint
 \ber
 \ito \mathrm{U}Q(\mathbf{U}-\mathbf{1})\ud x+\frac{ \mathbf{b}}{\lambda}=\mathbf{0}, \label{a21}
 \eer
 which implies
 \ber
  \ito \mathbf{U}^\tau Q(\mathbf{U}-\mathbf{1})\ud x+\frac{ \mathbf{1}^\tau\mathbf{b}}{\lambda}=0.\label{a22}
 \eer
 We may rewrite \eqref{a22} as
  \ber
  \ito \left(\mathbf{U}-\frac{\mathbf{1}}{2}\right)^\tau Q\left(\mathbf{U}-\frac{\mathbf{1}}{2}\right)\ud x&=&\frac{|\Omega|}{4}\mathbf{1}^\tau Q\mathbf{1}-\frac{ \mathbf{1}^\tau\mathbf{b}}{\lambda}\nm\\
  &=& \frac{|\Omega|}{4}\mathbf{1}^\tau P^{-1}  (K^\tau)^{-1}\mathbf{1}-\frac{ 4\pi\mathbf{1}^\tau P^{-1}(K^\tau)^{-1}\mathbf{N}}{\lambda},\label{a23}
 \eer
 where we have used the fact
 \ber
  (K^\tau)^{-1}\mathbf{1}=S^{-1}P^{-1}\mathbf{1}=R\mathbf{1}\label{a24}
 \eer
 and \eqref{a18}.

 Since the matrix  $Q$ is positive definite,  \eqref{a23}  gives  a necessary condition for the existence of solutions of \eqref{a15}:
 \berr
  \frac{|\Omega|}{4}\mathbf{1}^\tau P^{-1}  (K^\tau)^{-1}\mathbf{1}-\frac{ 4\pi\mathbf{1}^\tau P^{-1}(K^\tau)^{-1}\mathbf{N}}{\lambda}>0,
 \eerr
 which establishes (i) in Theorem \ref{th1}.

Now we show that the system  \eqref{a15} admits a variational formulation.  To see this we consider  the system \eqref{a15} in its  equivalent formulation, \eqref{a17}.

Now since the matrices $A$ and $Q$ defined in \eqref{a6a} are symmetric, we  see that the equations  \eqref{a17}  are the Euler--Lagrange equations of the functional
 \ber
  I(\mathbf{v})&=&\frac12\sum\limits_{i=1}^2\ito\partial_i\mathbf{v}^\tau A\partial_i\mathbf{v}\ud x+\frac\lambda2\ito(\mathbf{U}-\mathbf{1})^\tau Q(\mathbf{U}-\mathbf{1}) \ud x
  +\ito \frac{\mathbf{b}^\tau\mathbf{v}}{|\Omega|}\ud x. \label{a20}
 \eer
Here and in what follows we use the notation  \eqref{a6a},  \eqref{a13}, \eqref{a16} and  \eqref{a18}  without explicit reference.

We observe that the functional \eqref{a20} is not bounded from below. So we cannot conduct a direct minimization. To deal with  this problem,  we will find a critical point of the functional $I$ by using a constrained
minimization approach developed in \cite{CY1}, later refined by  \cite{nota}. Recently such a treatment  was extended by  \cite{hata} to solve the system  associated with some  general  $2\times2$  Cartan matrices.
To carry out  this constrained minimization, the main difficulty is how to resolve the constraints, which will be the focus of the next three sections.

\section{The  constraints}
\setcounter{equation}{0}
\setcounter{lemma}{0}
\setcounter{remark}{0}
In this section we identify a family of integral  constraints under which our variational functional will be minimized.

We start by decomposing the Sobolev space  $W^{1,2}(\Omega)$ into
$
  W^{1,2}(\Omega)=\mathbb{R}\oplus  \dot{W}^{1,2}(\Omega),
$
  where
\[\dot{W}^{1,2}(\Omega)\equiv\left\{w\in W^{1,2}(\Omega)\Bigg| \ito w\ud x=0\right\}\]
  is a closed subspace of $W^{1,2}(\Omega)$.
Then, for any $v_i\in W^{1, 2}(\Omega)$, we have
$v_i=c_i+w_i,   c_i \in \mathbb{R}, w_i\in  \dot{W}^{1,2}(\Omega), i=1, \dots, n. $
To save notation, in the sequel, we also interchangeably use $W^{1, 2}(\Omega), \dot{W}^{1, 2}(\Omega)$ to denote the spaces of both scalar and vector-valued  functions.
Hence, if $\mathbf{v}=\mathbf{w}+\mathbf{c}\in W^{1,2}(\Omega)$,  with $\mathbf{w}\equiv(w_1, \dots,  w_n)^\tau\in\dot{W}^{1,2}(\Omega)$ and $\mathbf{c}\equiv(c_1, \dots, c_n)^\tau\in \mathbb{R}^n$,   satisfies  the constraint  \eqref{a21}, we obtain
\ber
{\rm diag}\{\re^{c_1}, \dots, \re^{c_n}\} \tilde{Q}
\begin{pmatrix}\re^{c_1}\\\vdots \\ \re^{c_n}
\end{pmatrix} -P^{-1}R{\rm diag}\{a_1, \dots, a_n\} \begin{pmatrix}\re^{c_1}\\\vdots \\ \re^{c_n}
\end{pmatrix}+\frac{\mathbf{b}}{\lambda}=\mathbf{0},\label{a25a}
\eer
where
\ber
&&\tilde{Q}\equiv\tilde{Q}(\mathbf{w})\equiv R\tilde{S}R, \label{a25b}\\
&&\tilde{S}\equiv\begin{pmatrix}
\alpha_{11}a_{11}&-\alpha_{12}a_{12}& \cdots &\cdots&\cdots&-\alpha_{1n}a_{1n}\\
 \vdots& \vdots& \vdots &\vdots& \vdots&\vdots\\
 -\alpha_{i1}a_{i1}& -\alpha_{i2}a_{i2} &\cdots&\alpha_{ii}a_{ii}&\cdots&-\alpha_{in}a_{in}\\
 \vdots& \vdots& \vdots &\vdots&\vdots& \vdots\\
 -\alpha_{n1}a_{n1}& -\alpha_{n2}a_{n2} &\cdots&\cdots&\cdots &\alpha_{nn}a_{nn}
\end{pmatrix}\quad \label{a25b'}
\eer
 and  we adapt the notation
 \ber
a_i&\equiv& a_i(w_i)\equiv\ito\re^{u_i^0+w_i}\ud x,\label{a27}\\
 a_{ij}&\equiv& a_{ij}(w_i, w_j)\equiv\ito\re^{u_i^0+u_j^0+w_i+w_j}\ud x,
  \quad i, j=1, \dots, n. \label{a28}
\eer

Since the matrix  $S$ is positive definite,   from \eqref{a25b} and \eqref{a25b'}  we see that
\ber
 \tilde{Q}\quad \text{is positive definite.} \label{a25c}
\eer

Now the system \eqref{a25a} can be rewritten in its component form:
 \ber
 \re^{2c_i}R_i^2\alpha_{ii}a_{ii}-\re^{c_i}\left(\frac{R_ia_i}{P_i}+\sum\limits_{j\neq i}\re^{c_j}R_iR_j\alpha_{ij}a_{ij}\right)+\frac{b_i}{\lambda}=0, \, i=1, \dots, n.\label{a26}
 \eer

For any $\mathbf{w}\in \dot{W}^{1, 2}(\Omega)$, we see that  the equations \eqref{a26} with respect to $\mathbf{c}$   are solvable only if
 \ber
   \left(\frac{R_ia_i}{P_i}+\sum\limits_{j\neq i}\re^{c_j}R_iR_j\alpha_{ij}a_{ij}\right)^2&\ge& \frac{4R_i^2b_i\alpha_{ii}a_{ii}}{\lambda},  \quad i=1, \dots, n.\label{a29}
 \eer

In order to  ensure \eqref{a29},    it is sufficient to  take the following inequality-type constraints
 \ber
 \frac{ a_i^2}{a_{ii}}&\ge& \frac{4\alpha_{ii}P_i^2b_i}{\lambda}, \quad i=1, \dots, n.\label{a30}
 \eer

 Define the admissible set
\ber
\mathcal{A}&\equiv&\Big\{\mathbf{w} \big|\mathbf{w}\in \dot{W}^{1, 2}(\Omega) \text{ such that } \eqref{a30} \text{ is satisfied }\Big\}.\label{a30a}
\eer

Therefore,  for any $\mathbf{w}\in \mathcal{A}$,   we can obtain a solution of  \eqref{a26} by solving the system
  \ber
  \re^{c_i}&=&\frac{1}{2R_i^2\alpha_{ii}a_{ii}}
 \left\{\left(\frac{R_ia_i}{P_i}+\sum\limits_{j\neq i}\re^{c_j}R_iR_j\alpha_{ij}a_{ij}\right)\right.\nn\\
&&\left.+\sqrt{\left(\frac{R_ia_i}{P_i}+\sum\limits_{j\neq i}\re^{c_j}R_iR_j\alpha_{ij}a_{ij}\right)^2-\frac{4b_iR_i^2\alpha_{ii}a_{ii}}{\lambda}}\right\}
  \nn\\&\equiv& f_i(\re^{c_1},\dots,  \re^{c_n}),\quad i=1, \dots, n. \label{a31}
  \eer

 In the  next two  sections we aim at resolving  the constraints \eqref{a26} by solving \eqref{a31}.

\section{Resolving the $SU(4)$ constraints}
\setcounter{equation}{0}
\setcounter{lemma}{0}
\setcounter{remark}{0}

In this  section we present a direct/concrete method for resolving  the constraints \eqref{a26} when $K$ is the Cartan matrix of $SU(4)$.   Since in this case the coupling between the equations
enjoys some special properties, we will see that the constraints allow a ``squeeze-to-the-middle" solution process to be effectively carried out, which is of independent interest.

 For  $SU(4)$, the associated  Cartan matrix $K=(K_{ij})$ is given by
\be
K=\begin{pmatrix}
2&-1&0\\
-1&2&-1\\
0&-1&2
\end{pmatrix}.\label{z1}
\ee
Obviously,  now $K$ satisfies all the requirement in Theorem \ref{th1} with $P=I$.
Note that,   in this case,
 \ber
K^{-1}=\frac14\begin{pmatrix}
3&2&1\\
2&4&2\\
1&2&3\\
\end{pmatrix}, \label{z2}
\eer
 \ber
 R={\rm diag}\left\{\frac32,2,\frac32\right\}, \label{z3}
 \eer
\ber
\mathbf{b}=(b_1,b_2,b_3)^\tau=\frac14\begin{pmatrix}
3&2&1\\
2&4&2\\
1&2&3\\
\end{pmatrix}(N_1,N_2, N_3)^\tau. \label{z4}
\eer
Then the  constraints  \eqref{a26} are
 \ber
 &&\re^{2c_1}a_{11}-\re^{c_1}\left(\frac{a_1}{3}+\frac{2a_{12}}{3}\re^{c_2}\right)+\frac{2b_1}{9\lambda}=0,\label{a26'}\\
  &&\re^{2c_2}a_{22}-\re^{c_2}\left(\frac{a_2}{4}+\frac{3a_{12}}{8}\re^{c_1}+\frac{3a_{23}}{8}\re^{c_3}\right)+\frac{b_2}{8\lambda}=0,\label{a26''}\\
   &&\re^{2c_3}a_{33}-\re^{c_3}\left(\frac{a_3}{3}+\frac{2a_{23}}{3}\re^{c_2}\right)+\frac{2b_3}{9\lambda}=0.\label{a26'''}
 \eer

Furthermore, for any $\mathbf{w}\in \dot{W}^{1, 2}(\Omega)$,  to ensure the solvability of   the equations \eqref{a26'}--\eqref{a26'''} with respect to $\mathbf{c}$,
 the  required  inequality-type constraints  \eqref{a30} take the form
 \ber
 \frac{ a_i^2}{a_{ii}}&\ge& \frac{8b_i}{\lambda}, \quad i=1,2, 3. \label{a30'}
 \eer
The admissible set $\mathcal{A}$ reads
\ber
\mathcal{A}&\equiv&\Big\{\mathbf{w} \big|\mathbf{w}\in \dot{W}^{1, 2}(\Omega)  \text{ such that } \eqref{a30'} \text{ is satisfied}\Big\}.\label{a30a'}
\eer

Hence,  for any $\mathbf{w}\in \mathcal{A}$,  to get a solution of    \eqref{a26'}--\eqref{a26'''}, it suffices to solve
   \ber
 \re^{c_1}&=&\frac{\frac{a_1}{3}+\frac{2a_{12}}{3}\re^{c_2}+\sqrt{\left(\frac{a_1}{3}+\frac{2a_{12}}{3}\re^{c_2}\right)^2-\frac{8b_1a_{22}}{9\lambda}}}{2a_{11}}\equiv f_1(\re^{c_2}), \label{z5}\\
  \re^{c_2}&=&\frac{ \frac{a_2}{4}+\frac{3a_{12}}{8}\re^{c_1}+\frac{3a_{23}}{8}\re^{c_3}+\sqrt{\left(\frac{a_2}{4}+\frac{3a_{12}}{8}\re^{c_1}+\frac{3a_{23}}{8}\re^{c_3}\right)^2-\frac{ b_2a_{22}}{2\lambda}}}{2a_{22}}\equiv f_2(\re^{c_1}, \, \re^{c_3}),\label{z6}\\
  \re^{c_3}&=&\frac{\frac{a_3}{3}+\frac{2a_{23}}{3}\re^{c_2}+\sqrt{\left(\frac{a_3}{3}+\frac{2a_{23}}{3}\re^{c_2}\right)^2-\frac{8b_3a_{33}}{9\lambda}}}{2a_{33}}\equiv f_3(\re^{c_2}).\label{z7}
 \eer
To solve these equations, we simply need to find a positive zero of the function
   \be
    F(t)\equiv t-f_2\big(f_1(t), f_3(t)\big), \quad t\in [0, \infty). \label{z8}
   \ee

We have
 \begin{proposition}\label{prop1z}
 For any $\mathbf{w}\in   \mathcal{A}$,  there exists a unique positive  solution  $t_0$ for the equation
  \[F(t)\equiv t-f_2\big(f_1(t), f_3(t)\big)=0.\]
 \end{proposition}

In view of this proposition we see that, for any $\mathbf{w}\in  \mathcal{A}$,  the system \eqref{z5}--\eqref{z7} with respect to $\mathbf{c}$ admits a unique solution. However, as shown in the next section  for the general case, there is no guarantee for the uniqueness of a solution to the general system \eqref{a31}.

 {\bf Proof. }   \quad  For any $\mathbf{w}\in\mathcal{A}$,  we see from \eqref{z5}--\eqref{z7} that
 \be
 f_i(t)>0, \quad \forall\, t\ge0, \, i=1, 3,  \quad f_2(t, s)>0,\, \forall\,t, s\ge0, \label{z9}
 \ee
which imply
  \[F(0)=-f_2(f_1(0), f_3(0))<0. \]
    A direct computation gives
    \ber
    \lim\limits_{t\to\infty}\frac{f_1(t)}{t}&=&\frac{2a_{12}}{3a_{11}},  \label{z10}\\
     \lim\limits_{t\to\infty}\frac{f_2(t, t)}{t}&=&\frac{3(a_{12}+a_{23})}{8a_{22}}, \label{z11} \\
    \lim\limits_{t\to\infty}\frac{f_3(t)}{t}&=&\frac{2a_{23}}{3a_{33}}.\label{z12}
    \eer
Then, using the above limits and the H\"{o}lder inequality, we have
   \ber
  \lim\limits_{t\to\infty}\frac{F(t)}{t}&=&1-\lim\limits_{t\to\infty}\frac{f_2\big(f_1(t), f_3(t)\big)}{t}\nn \\
  &=&1- \frac{a_{12}^2}{4a_{11}a_{22}}-\frac{a_{23}^2}{4a_{22}a_{33}}\nn\\
   &\ge& 1-\frac14-\frac14>0. \label{z13}
   \eer
 Then we see that  \be\lim\limits_{t\to\infty}F(t)=\infty.\label{z14}\ee

Since we have   $F(0)<0$,   then  the function  $F(\cdot)$  admits at least one  zero  $t_0\in (0, \infty)$.

 Now we  prove that the  zero of $F(\cdot)$ is also unique.
  In fact we easily check that
 \ber
  \frac{\ud f_1(t)}{\ud t}&=& \frac{\frac{2a_{12}}{3}f_1(t) }{\sqrt{\left(\frac{a_1}{3}+\frac{2a_{12}}{3}t\right)^2-\frac{8b_1a_{11}}{9\lambda}}}, \label{z15}\\
  \frac{\partial f_2(t, s)}{\partial t}&=& \frac{\frac{3a_{12}}{8}f_2(t, s)}{\sqrt{\left(\frac{a_2}{4}+\frac{3a_{12}}{8}t+\frac{3a_{23}}{8}s\right)^2-\frac{b_2a_{22}}{2\lambda}}},\label{z16}\\
  \frac{\partial f_2(t, s)}{\partial s}&=& \frac{\frac{3a_{23}}{8}f_2(t, s)}{\sqrt{\left(\frac{a_2}{4}+\frac{3a_{12}}{8}t+\frac{3a_{23}}{8}s\right)^2-\frac{b_2a_{22}}{2\lambda}}},\label{z17}\\
  \frac{\ud f_3(t)}{\ud t}&=& \frac{\frac{2a_{23}}{3}f_3(t)}{\sqrt{\left(\frac{a_3}{3}+\frac{2a_{23}}{3}t\right)^2-\frac{8b_3a_{33}}{9\lambda}}},\label{z18}
 \eer
  which are all positive.  Thus,  the functions $f_i(t)$, $(i=1, 3)$ are  strictly increasing for  all  $t>0$.

Then from \eqref{z15}--\eqref{z18}, \eqref{z5}--\eqref{z7}  and the constraints \eqref{a30'}  we obtain
 \berr
 \frac{\ud F(t)}{\ud t}&=&1-\frac{\frac14f_2\big(f_1(t), f_3(t)\big)}{\sqrt{\left(\frac{a_2}{4}+\frac{3a_{12}}{8}f_1(t)+\frac{3a_{23}}{8}f_3(t)\right)^2-\frac{b_2a_{22}}{2\lambda}}}\times\\
 &&\times\left[\frac{a_{12}^2f_1(t)}{\sqrt{\left(\frac{a_1}{3}+\frac{2a_{12}}{3}t\right)^2-\frac{8b_1a_{11}}{9\lambda}}}+\frac{a_{23}^2f_3(t)}{\sqrt{\left(\frac{a_3}{3}+\frac{2a_{23}}{3}t\right)^2-\frac{8b_3a_{33}}{9\lambda}}}\right]\\
 &>&1-\frac{\frac38f_2\big(f_1(t), f_3(t)\big)}{\sqrt{\left(\frac{a_2}{4}+\frac{3a_{12}}{8}f_1(t)+\frac{3a_{23}}{8}f_3(t)\right)^2 -\frac{b_2a_{22}}{2\lambda}}}\frac{a_{12}f_1(t)+a_{23}f_3(t)}{t}\\
 &>&1-\frac{\frac38f_2\big(f_1(t), f_3(t)\big)}{\frac{3a_{12}}{8}f_1(t)+\frac{3a_{23}}{8}f_3(t)}\frac{a_{12}f_1(t)+a_{23}f_3(t)}{t}\\
 &=&1-\frac{f_2\big(f_1(t), f_3(t)\big)}{t}=\frac{F(t)}{t},\quad t>0,
 \eerr
   so the uniqueness of the zero of  $F(t)$ over $[0, \infty)$ follows from the monotonicity of $F(t)/t$ for $t>0$ and
the proof of the proposition is complete.

Using  Proposition \ref{prop1z},   for any  $\mathbf{w}\in\mathcal{A}$,   we see that    the equations \eqref{a26'}--\eqref{a26'''}  with respect to $ \mathbf{c}$
 admit a unique solution   $\mathbf{c}(\mathbf{w})=\big(c_1(\mathbf{w}), c_2(\mathbf{w}), c_3(\mathbf{w})\big)^\tau$ determined  by
 \eqref{z5}--\eqref{z7},  such that  $ \mathbf{v}=(v_1, v_2, v_3)^\tau$ defined by
 \be v_i=w_i+c_i(\mathbf{w}),  \quad i=1, 2, 3, \label{c24a}\ee
 satisfies  the constraints  \eqref{a21} for the $SU(4)$ case.

\section{Solving the constraints in general}
\setcounter{equation}{0}
\setcounter{lemma}{0}
\setcounter{remark}{0}
\setcounter{proposition}{0}
In this  section we carry out a new way  to resolve  the constraints \eqref{a26} by solving  \eqref{a31} for the general case via a topological-degree-theory argument.
To  this end, we consider  the system
 \ber
  \mathbf{F}(\mathbf{t})\equiv \mathbf{t}-\mathbf{f}(\mathbf{t})=\mathbf{0}, \quad \mathbf{t}\equiv (t_1, \dots, t_n)^\tau \in \mathbb{R}^n_+,\label{a32}
 \eer
where
$
 \mathbb{R}^n_+\equiv(\mathbb{R}_+)^n,\mathbf{f}(\mathbf{t})\equiv (f_1(\mathbf{t}), \dots, f_n(\mathbf{t}))^\tau.%\label{a33}
$
We have

\begin{proposition}\label{prop1}
 For any $\mathbf{w}\in\mathcal{A}$,  the system \eqref{a32} admits a solution $\mathbf{t}\in (0, \infty)^n$.
\end{proposition}

{\bf Proof. }
To conduct a  degree-theory argument,  we  deform the system \eqref{a32} as
 \ber
  \mathbf{F}(\epsilon, \mathbf{t})\equiv \mathbf{t}-\mathbf{f}(\epsilon, \mathbf{t})=\mathbf{0}, \quad \mathbf{t}\in \mathbb{R}^n_+,\quad \epsilon\in [0, 1], \label{a34}
 \eer
where
 \ber
 \mathbf{f}(\epsilon, \mathbf{t})&\equiv&(f_1(\epsilon, \mathbf{t}), \dots, f_n(\epsilon, \mathbf{t}))^\tau, \label{a35}\\
   f_i(\epsilon, \mathbf{t})&\equiv&\frac{1}{2R_i^2\alpha_{ii}a_{ii}}
 \left\{\left(\frac{R_ia_i}{P_i}+\sum\limits_{j\neq i}t_jR_iR_j\alpha_{ij}a_{ij}\right)\right.\nn\\
&&\left.+\sqrt{\left(\frac{R_ia_i}{P_i}+\sum\limits_{j\neq i}t_jR_iR_j\alpha_{ij}a_{ij}\right)^2-\frac{4\epsilon b_iR_i^2\alpha_{ii}a_{ii}}{\lambda}}\right\},
  \nn\\&& \quad i=1, \dots, n. \label{a36}
\eer
Then, we see that, to solve the  system  \eqref{a32}, we need to find a solution of
 \ber
  \mathbf{F}(1, \mathbf{t})=\mathbf{0}, \quad \mathbf{t}\in \mathbb{R}^n_+. \label{a36a}
 \eer

To facilitate our statement, we use the convention that we write
 \[
(\alpha_1,\dots, \alpha_n)^\tau<(\le )\,(\beta_1,\dots, \beta_n)^\tau \mbox{ if }\alpha_i<(\le)\,\beta_i, i=1,\dots, n,
\]
and we use the same convention for matrices.

To proceed, we  establish the following key {\em a priori} estimates.

\begin{lemma}\label{lem0}
  For any $\mathbf{w}\in \mathcal{A}$ and  $\epsilon\in[0, 1]$,
every  solution $\mathbf{t}$ of \eqref{a34} satisfies
  \ber
    0<a_it_i\le |\Omega|,\quad i=1, \dots, n,\label{a37}\\
     0<t_i\le 1, \quad \quad i=1, \dots, n.\label{a37a}
\eer
\end{lemma}

By virtue of Lemma  \ref{lem0} we immediately obtain the following.
\begin{corollary}
  For any $\mathbf{w}\in \mathcal{A}$, every  solution $(c_1, \dots, c_n)$ of \eqref{a31} satisfies
  \ber
   a_i\re^{c_i}\le|\Omega|,\quad i=1, \dots, n,\label{a37b}\\
     \re^{c_i}\le1, \quad \quad i=1, \dots, n.\label{a37c}
\eer
\end{corollary}

{\bf Proof of Lemma \ref{lem0}.}
 For any $\mathbf{w}\in \mathcal{A}$ and  $\epsilon\in[0, 1]$, let $\mathbf{t}$ be a solution of \eqref{a34}.
 By \eqref{a34}--\eqref{a36} we readily   get   the left-hand sides of \eqref{a37}--\eqref{a37a}.

From \eqref{a34} it is straightforward  to see that
 \ber
   t_i=f_i(\epsilon, \mathbf{t})\le \frac{\frac{R_ia_i}{P_i}+\sum\limits_{j\neq i}t_jR_iR_j\alpha_{ij}a_{ij}}{R_i^2\alpha_{ii}a_{ii}}, \quad i=1, \dots, n,\label{a38}
 \eer
 which can be rewritten as
 \ber
 R_i^2\alpha_{ii}a_{ii}t_i-\sum\limits_{j\neq i}t_jR_iR_j\alpha_{ij}a_{ij}\le\frac{R_ia_i}{P_i}, \quad i=1, \dots, n.\label{a38a}
\eer

Noting the expression of $\tilde{Q}$,   we rewrite  \eqref{a38a}   in a vector form
 \ber
\tilde{Q} \mathbf{t} \le P^{-1}R\mathbf{a},\label{a39}
\eer
where
$
  \mathbf{a}\equiv (a_1,\dots,a_n)^\tau.
$
By H\"{o}lder's  inequality we   obtain
 \ber
   a_{ij}^2\le a_{ii}a_{jj}, \quad a_i\le |\Omega|^{\frac12}a_{ii}^{\frac12}, \quad i,j=1,\dots, n.\label{a01}
 \eer

Therefore,  noting that $t_i>0,\, i=1,\dots, n$, the expression of $\tilde{Q}$,  and   \eqref{a01},  we  arrive at
  \ber
 {\rm{diag}}\left\{a_{11}^{\frac12},\dots,a_{nn}^{\frac12}\right\}Q {\rm{diag}}\left\{a_{11}^{\frac12},\dots,a_{nn}^{\frac12}\right\}\mathbf{t} \le \tilde{Q}\mathbf{t}. \label{a47a}
\eer

 On the other hand, by  \eqref{a5z4} and the expression of $Q$ in \eqref{a6a}  we see  that
 \ber
  \text{all the entries of }   Q^{-1} \text {are positive. }\label{a05}
  \eer
 Therefore, by  \eqref{a39},  \eqref{a47a}, and \eqref{a05},  we get
\ber
 \mathbf{t}&\le& {\rm{diag}}\left\{a_{11}^{-\frac12},\dots,a_{nn}^{-\frac12}\right\}Q^{-1} {\rm{diag}}\left\{a_{11}^{-\frac12},\dots,a_{nn}^{-\frac12}\right\}P^{-1}R\mathbf{a}\nn\\
  &=&{\rm{diag}}\left\{a_{11}^{-\frac12},\dots,a_{nn}^{-\frac12}\right\}Q^{-1} {\rm{diag}}\left\{a_1a_{11}^{-\frac12},\dots,a_na_{nn}^{-\frac12}\right\}P^{-1}R\mathbf{1}. \label{a48a}
\eer

From   \eqref{a01}, \eqref{a05}, and  \eqref{a48a}  we infer that
\ber
 &&{\rm{diag}}\{a_1,\dots,a_n\}\mathbf{t}\nm\\
 &&\le{\rm{diag}}\left\{a_1a_{11}^{-\frac12},\dots,a_na_{nn}^{-\frac12}\right\}Q^{-1} {\rm{diag}}\left\{a_1a_{11}^{-\frac12},\dots,a_na_{nn}^{-\frac12}\right\}P^{-1}R\mathbf{1}\nn\\
  &&\le|\Omega|Q^{-1}P^{-1}R\mathbf{1}\nn\\
  &&=|\Omega|\mathbf{1}, \label{a48b}
\eer
where \eqref{a24}  is  also used . Then the right-hand sides of   \eqref{a37} follow from \eqref{a48b}.
By Jensen's inequality, we have $a_i\ge |\Omega|, i=1, \dots, n$,  which together with \eqref{a48b}  imply the right-hand sides of \eqref{a37a}.   The proof of Lemma \ref{lem0} is complete.

By the definition of $\mathbf{F}(\epsilon, \mathbf{t})$, for any $\epsilon\in[0, 1]$, it is easy to see that  $\mathbf{F}(\epsilon, \mathbf{t})$ is a smooth function from
$\mathbb{R}^n_+$ into $\mathbb{R}^n$.

Let us define
 \be
\tilde{\Omega}\equiv (0, r_0)^n,\label{a39b}
\ee
 where $r_0>1$ is a constant.

Then  by  \eqref{a37a}  we see that, for every $\mathbf{w}\in\mathcal{A}$ and $\vep\in[0, 1]$,  $\mathbf{F}(\epsilon, \mathbf{t})$ has no zero on the boundary of $\tilde{\Omega}$. Consequently,  the Brouwer degree
 \be
 \deg(\mathbf{F}(\epsilon, \mathbf{t}), \tilde{\Omega}, \mathbf{0})\label{a39c}
\ee
is well defined.

To prove Proposition \ref{prop1}, it is  suffficient to show that
  \be
 \deg(\mathbf{F}(1, \mathbf{t}), \tilde{\Omega}, \mathbf{0})\neq0.\label{a39d}
\ee
Since  $\mathbf{F}(\epsilon, \mathbf{t})$ is smooth, by homotopy invariance \cite{Nire},  we have
  \be
 \deg(\mathbf{F}(1, \mathbf{t}), \tilde{\Omega}, \mathbf{0})=\deg(\mathbf{F}(0, \mathbf{t}), \tilde{\Omega}, \mathbf{0}).\label{a39e}
\ee
Now we only need to calculate the degree $\deg(\mathbf{F}(0, \mathbf{t}), \tilde{\Omega}, \mathbf{0})$.

Note for any    $\mathbf{w}\in \mathcal{A}$ the system
 \be
 \mathbf{F}(0, \mathbf{t})=\mathbf{0} \label{a39f}
\ee
is reduced into
 \ber
 t_i-\frac{\frac{R_ia_i}{P_i}+\sum\limits_{j\neq i}t_jR_iR_j\alpha_{ij}a_{ij}}{R_i^2\alpha_{ii}a_{ii}}=0, \quad  i=1, \dots, n.\label{a39g}
 \eer
By the definition of $\tilde{Q}$  we rewrite the system \eqref{a39g} equivalently  in a vector form
\ber
 \tilde{Q}\mathbf{t}=P^{-1}R\mathbf{a}.\label{a39h}
\eer

In view of \eqref{a25c} the matrix  $\tilde{Q}$   is of course    invertible. Then   we see that  the system \eqref{a39h}, i.e., \eqref{a39f},  has a unique solution
 \berr
 \mathbf{t}=\tilde{Q}^{-1}P^{-1}R\mathbf{a},
\eerr
which, belonging  to $\tilde{\Omega}$, is not  a boundary point  of $\tilde{\Omega}$  by \eqref{a37a}.

 Noting  \eqref{a25c},  the determinant of $\tilde{Q}$ is positive, which implies that  the Jacobian of $\mathbf{F}(0, \mathbf{t})$ is positive everywhere.
Therefore, by the definition of the Brouwer degree,  we have
$
 \deg(\mathbf{F}(0, \mathbf{t}), \tilde{\Omega}, \mathbf{0})=1,
$
which implies
$
 \deg(\mathbf{F}(1, \mathbf{t}), \tilde{\Omega}, \mathbf{0})=1.
$
Then the proof of Proposition \ref{prop1} is complete.

By Proposition \ref{prop1},  we see that for any $\mathbf{w}\in \mathcal{A}$, the constraint equations \eqref{a26} admit a
solution
 $\mathbf{c}(\mathbf{w})=(c_1(\mathbf{w}), \dots, c_n(\mathbf{w}))^\tau$
determined by \eqref{a31}, such that
$\mathbf{v}=\mathbf{w}+\mathbf{c}(\mathbf{w})=(w_1+c_1(\mathbf{w}), \dots, w_n+c_n(\mathbf{w}))^\tau$
satisfies the constraints  \eqref{a21}.

\section{Constrained minimization}
\setcounter{equation}{0}
\setcounter{lemma}{0}
 In this section we solve the equation \eqref{a14} by finding a critical point of the functional $I$ via  a constrained minimization procedure. To do this, we consider the
constrained functional
\ber
 J(\mathbf{w})\equiv I(\mathbf{w}+\mathbf{c}(\mathbf{w})),\quad  \mathbf{w}\in \mathcal{A}, \label{a40}
\eer
where $\mathbf{c}(\mathbf{w})$ is the solution of   the constraint equations \eqref{a26}  determined by  Proposition \ref{prop1}.

 Since, for every $\mathbf{w}\in\mathcal{A}$, $\mathbf{v}=\mathbf{w}+\mathbf{c}(\mathbf{w})$ satisfies the
constraints \eqref{a21},  we have
 \ber
 \ito (\mathbf{U}-\mathbf{1})^\tau Q(\mathbf{U}-\mathbf{1}) \ud x &=& \ito \mathbf{1}^\tau Q(\mathbf{1}-\mathbf{U})\ud x
 -\frac{\mathbf{1}^\tau \mathbf{b}}{\lambda} \nn\\
 &=&\ito \mathbf{1}^\tau P^{-1}R(\mathbf{1}-\mathbf{U})\ud x -\frac{\mathbf{1}^\tau \mathbf{b}}{\lambda}.\label{a41}
 \eer
Then  the functional $J$  can be expressed as
 \ber
  J(\mathbf{w})& =&\frac12\sum\limits_{i=1}^2\ito \partial_i\mathbf{w}^\tau A\partial_i\mathbf{w}\ud x
 + \frac\lambda2\mathbf{1}^\tau P^{-1}R\ito (\mathbf{1}-\mathbf{U})\ud x +\mathbf{b}^\tau\mathbf{c}-\frac{\mathbf{1}^\tau \mathbf{b}}{2}\nn\\
 &=&\frac12\sum\limits_{i=1}^2\ito \partial_i\mathbf{w}^\tau A\partial_i\mathbf{w}\ud x
 +\frac{\lambda}{2}\sum\limits_{i=1}^n\frac{R_i}{P_i}\ito (1-\re^{c_i}\re^{u_i^0+w_i})\ud x \nn\\
 &&+\sum\limits_{i=1}^nb_ic_i-\frac12\sum\limits_{i=1}^nb_i, \label{a42}
 \eer
where and in the following we use the notation \eqref{a6a}, \eqref{a18}.

We easily see that the functional $J$ is Fr\'{e}chet differentiable in $\mathcal{A}$. We aim to  show that the functional $J$
admits a minimizer, say $\mathbf{w}$, in the interior of $\mathcal{A}$. Then $\mathbf{v}=\mathbf{w}+\mathbf{c}(\mathbf{w})$ is a critical point of the functional $I$.

To proceed further,  we need   the following    inequalities.
\begin{lemma}\label{lem2}
 For any $\mathbf{w}\in \mathcal{A}$ and $s\in (0, 1)$, there hold the inequalities
 \ber
 \ito \re^{u_i^0+w_i}\ud x\le \left(\frac{\lambda}{4P_i^2b_i\alpha_{ii}}\right)^{\frac{1-s}{s}}\left(\ito\re^{su_i^0+sw_i}\ud x\right)^{\frac{1}{s}}
  ,\quad i=1, \dots, n.\label{a50}\eer
\end{lemma}

Such type of  inequalities  were first established in \cite{nota1}.

For our purposes, we will  also need  the Moser--Trudinger inequality \cite{font}
 \ber
 \ito \re^w\ud x\le C\exp\left(\frac{1}{16\pi}\|\nabla w\|_2^2\right), \quad \forall w\in \dot{W}^{1,2}(\Omega), \label{a51}
\eer
where $C>0$ is a constant.

 Let $\alpha_0$ and $\beta_0$ be the smallest eigenvalues of $A$ and $Q$ defined by  \eqref{a6a}, respectively.

\begin{lemma}\label{lem3}
 For  every $\mathbf{w}\in \mathcal{A}$,  the functional $J$ satisfies
 \ber
 J(\mathbf{w}) \ge \frac{\alpha_0}{4}\sum\limits_{i=1}^n\|\nabla w_i\|_2^2-C(\ln\lambda+1), \label{a52}
\eer
 where   $C>0$  is a constant independent of $\lambda$.
\end{lemma}

{\bf Proof.}   By \eqref{a6a}  and the definition of $J$ we see that
  \ber
  J(\mathbf{w}) \ge \frac{\alpha_0}{2}\sum\limits_{i=1}^n\|\nabla w_i\|_2^2+\sum\limits_{i=1}^nb_ic_i.\label{a53}
\eer

Noting \eqref{a31} we  obtain
 \ber
 \re^{c_i}\ge \frac{a_i}{2P_iR_i\alpha_{ii}a_{ii}},\quad i=1, \dots, n,\label{a54}
\eer
which together with \eqref{a30} yield
  \ber
 \re^{c_i}\ge \frac{2P_ib_i}{\lambda R_ia_i}=\frac{2P_ib_i}{\lambda R_i\ito\re^{u_i^0+w_i}\ud x},\quad i=1, \dots, n.\label{a55}
\eer
Then we have
 \ber
 c_i\ge \ln\frac{2P_ib_i}{R_i}-\ln\lambda-\ln\ito\re^{u_i^0+w_i}\ud x,\quad i=1, \dots, n.\label{a56}
\eer

Using  Lemma \ref{lem2} and the Moser--Trudinger inequality \eqref{a51} we arrive at
\ber
\ln\ito\re^{u_i^0+w_i}\ud x&\le& \frac{1-s}{s}\left\{\ln\lambda-\ln\left(4P_i^2b_i\alpha_{ii}\right)\right\}
 +\frac1s\ln\ito\re^{su_i^0+sw_i}\ud x\nn\\
  &\le& \frac{s}{16\pi}\|\nabla w_i\|_2^2+\frac{1-s}{s}\left\{\ln\lambda-\ln\left(4P_i^2b_i\alpha_{ii}\right)\right\}
 \nn\\&&+\frac{\ln C}{s}+\max\limits_{\Omega} u_i^0, \quad i=1, \dots, n. \label{a57}
\eer

Combining \eqref{a53}, \eqref{a56}, and \eqref{a57}, we have
\ber
 J(\mathbf{w})&\ge& \left(\frac{\alpha_0}{2}-\frac{s\max\limits_{1\le i \le n}\{b_i\}}{16\pi}\right)\sum\limits_{i=1}^n\|\nabla w_i\|_2^2
    -\frac1s\sum\limits_{i=1}^nb_i\left\{\ln\lambda-\ln\left(4P_i^2b_i\alpha_{ii}\right)+\ln C\right\}
 \nn\\&&-\sum\limits_{i=1}^nb_i\left\{\ln\left(2R_iP_ib_i\alpha_{ii}\right)+\max\limits_{\Omega} u_i^0\right\}.\label{a58}
\eer
Therefore, taking $s$ suitably small in \eqref{a58} we obtain the lemma.

Using Lemma \ref{lem3},  we see that the functional $J$ is bounded from below and coercive in $\mathcal{A}$.
Then noting that   $J$ is weakly lower semicontinuous in $\mathcal{A}$,   we conclude that $J$  admits a minimizer in
$\mathcal{A}$.  In the following  we   establish   some estimates  as in   \cite{CY1,nota}  to  show that this minimizer belongs to the interior of $\mathcal{A}$ when $\lambda$ is
sufficiently large.

\begin{lemma}\label{lem4}
There exists a constant $C>0$ independent of $\lambda$ such that
 \ber
   \inf\limits_{\mathbf{w}\in\partial\mathcal{A}} J(\mathbf{w}) \ge \frac{|\Omega|\lambda}{2}\min\limits_{1\le i\le n}\left\{\frac{R_i}{P_i}\right\}-C(1+\ln\lambda+\sqrt{\lambda}).\label{a59}
 \eer
\end{lemma}

{\bf Proof. } By the definition of $\mathcal{A}$, we see that at least one of the following equalities
  \ber
    \frac{a_i^2}{a_{ii}} =\frac{4\alpha_{ii}P_i^2b_i}{\lambda}, \quad i=1, \dots, n \label{a60}
 \eer
must hold  on the boundary $\partial\mathcal{A}$.

If the case  $i=1$ in \eqref{a60}  holds,  by  \eqref{a48b} we obtain
 \ber
a_1\re^{c_1}&\le&\frac{R_1}{P_1}(Q^{-1})_{11}a_1^2a_{11}^{-1}+\sum\limits_{j=2}^n \frac{R_j}{P_j}(Q^{-1})_{1j}a_1a_ja_{11}^{-\frac12}a_{jj}^{-\frac12}
\nn\\
&\le&\frac{R_1}{P_1}(Q^{-1})_{11}a_1^2a_{11}^{-1}+ |\Omega|^{\frac12}\sum\limits_{j=2}^n \frac{R_j}{P_j}(Q^{-1})_{1j}a_1a_{11}^{-\frac12}
\nn\\
&=&(Q^{-1})_{11}\frac{4P_1R_1b_1\alpha_{11}}{\lambda}+\frac{2P_1\sum\limits_{j=2}^n \frac{R_j}{P_j}(Q^{-1})_{1j}}{\sqrt{\lambda}}\sqrt{b_1|\Omega|\alpha_{11}}.\label{a61}
\eer

 If other cases happen,   similar estimates as \eqref{a61} can be established.

Using \eqref{a37b} and \eqref{a61} leads to
\ber
&&\frac{\lambda}{2}\sum\limits_{i=1}^n\frac{R_i}{P_i}\ito (1-\re^{c_i}\re^{u_i^0+w_i})\ud x\nn\\
&&\ge\frac{|\Omega|\lambda R_1}{2P_1}-2(Q^{-1})_{11}P_1R_1b_1\alpha_{11}-P_1\sum\limits_{j=2}^n \frac{R_j}{P_j}(Q^{-1})_{1j}\sqrt{b_1\lambda|\Omega|\alpha_{11}}.\label{a62}
\eer

 By \eqref{a62} and estimating $c_i$ $ (i=1, \dots, n)$ as that in Lemma \ref{lem3}, we get the lemma.

  Now we need to choose some suitable   functions in the interior of $\mathcal{A}$  to estimate the value of the functional $J$ in $\mathcal{A}$.

Recall that in \cite{taran96} Tarantello proved that, for $\mu$ sufficiently large,  the problems
 \ber
 \Delta v=\mu \re^{u_i^0+v}(\re^{u_i^0+v}-1)+\frac{4\pi N_i}{|\Omega|}, \quad i=1, \dots, n,\label{a63}
\eer
 admit solutions $v_i^\mu (i=1, \dots, n) $, such that   $u_i^0+v_i^\mu<0$ in $\Omega$,
 $c_i^\mu=\frac{1}{|\Omega|}\ito v_i^\mu\ud x\to 0$ and $w_i^\mu=v_i^\mu-c_i^\mu\to -u_i^0$ pointwise
as $\mu\to\infty$, $i=1, \dots, n$. Then it follows  that
 \ber
 \lim\limits_{\mu\to\infty}\ito\re^{u_i^0+w_i^\mu}\ud x=|\Omega|,\quad
\lim\limits_{\mu\to\infty}\ito\re^{u_i^0+u_j^0+w_i^\mu+w_j^\mu}\ud x=|\Omega|, \quad i, j=1, \dots, n.\label{a64}
 \eer
Consequently, in view of the expression of $\tilde{Q}$,  we have
 \ber
  \lim\limits_{\mu\to\infty}\tilde{Q}(\mathbf{w}^\mu)=|\Omega|Q.\label{a65}
 \eer
 Noting \eqref{a25c},   we know that  $\tilde{Q}(\mathbf{w}^\mu)$ is invertible, which together with \eqref{a65} yield
 \ber
  \lim\limits_{\mu\to\infty}\tilde{Q}^{-1}(\mathbf{w}^\mu)=\frac{1}{|\Omega|}Q^{-1}.\label{a66}
 \eer

Therefore, by the definition of $\mathcal{A}$, \eqref{a64} and \eqref{a66},   we conclude that, for a fixed  $\lambda_0'>0$
sufficiently large and  any $\vep\in(0, 1)$, there exists a $\mu_\vep\gg1$, such that
 \ber
 \mathbf{w}^{\mu_\vep}=(w_1^{\mu_\vep}, \dots, w_n^{\mu_\vep})^\tau \in {\rm int}\mathcal{A},  \label{a67}
\eer
for every $\lambda>\lambda_0'$, and
 \ber
  &&a_{ij}(w_i^{\mu_\vep}, w_j^{\mu_\vep})<(1+\vep)|\Omega| < 2|\Omega|, \quad i,j=1, \dots, n,\label{a68}\\
  &&\frac{(1-\vep)}{|\Omega|}Q^{-1}<\tilde{Q}^{-1}(\mathbf{w}^{\mu_\vep})< \frac{(1+\vep)}{|\Omega|}Q^{-1}<\frac{2}{|\Omega|}Q^{-1}. \label{a69}
 \eer

At this point we can establish the following comparison result .

\begin{lemma} \label{lem5}
  For  $\mathbf{w}^{\mu_\vep}$ given by \eqref{a67},  we have
 \ber
 J(\mathbf{w}^{\mu_\vep})-\inf\limits_{\mathbf{w}\in \partial \mathcal{A}}J(\mathbf{w})<-1,\label{a70}
 \eer
when $\lambda$ is large enough.
\end{lemma}

{\bf Proof. } Since $\mathbf{w}^{\mu_\vep}\in {\rm int}\mathcal{A}$, by \eqref{a31} and the Jensen inequality, we arrive at the estimates
  \ber
 \re^{c_i(\mathbf{w}^{\mu_\vep})}&=&\frac{\frac{R_ia_i}{P_i}+\sum\limits_{j\neq i}\re^{c_j(\mathbf{w}^{\mu_\vep})}R_iR_j\alpha_{ij}a_{ij}}{2R_i^2\alpha_{ii}a_{ii}}
 \nn\\&&\times\left(1+\sqrt{1-\frac{4b_iR_i^2\alpha_{ii}a_{ii}}{\lambda\left(\frac{R_ia_i}{P_i}+\sum\limits_{j\neq i}\re^{c_j}R_iR_j\alpha_{ij}a_{ij}\right)^2}}\right)
 \nn\\
&\ge&\frac{\frac{R_ia_i}{P_i}+\sum\limits_{j\neq i}\re^{c_j(\mathbf{w}^{\mu_\vep})}R_iR_j\alpha_{ij}a_{ij}}{R_i^2\alpha_{ii}a_{ii}}
-\frac{2b_i}{\lambda\left(\frac{R_ia_i}{P_i}+\sum\limits_{j\neq i}\re^{c_j}R_iR_j\alpha_{ij}a_{ij}\right) }\nn\\
&\ge&\frac{\frac{R_i|\Omega|}{P_i}+\sum\limits_{j\neq i}\re^{c_j(\mathbf{w}^{\mu_\vep})}R_iR_j\alpha_{ij}a_{ij}}{R_i^2\alpha_{ii}a_{ii}}
-\frac{2P_ib_i}{\lambda|\Omega|R_i}, \quad i=1, \dots, n.\label{a71}
\eer
Here and to the end of this section  we understand that
\ber
a_i=a_i(w_i^{\mu_\vep}), \quad a_{ij}=a_{ij}(w_i^{\mu_\vep}, w_j^{\mu_\vep}), \quad i, j=1,\dots, n.
\eer

 Then it follows  from    \eqref{a71}  and  \eqref{a68} that
\ber
 R_i^2\alpha_{ii}a_{ii}\re^{c_i(\mathbf{w}^{\mu_\vep})}-\sum\limits_{j\neq i}\re^{c_j(\mathbf{w}^{\mu_\vep})}R_iR_j\alpha_{ij}a_{ij}
 &\ge&\frac{R_i|\Omega|}{P_i}-\frac{2P_iR_ib_i}{\lambda|\Omega|}\alpha_{ii}a_{ii}\nn\\
   &\ge& \frac{R_i|\Omega|}{P_i}-\frac{4\alpha_{ii}P_iR_ib_i}{\lambda} , \quad i=1, \dots, n,\label{a72}
\eer
which can be expressed  in a vector form:
 \ber
 \tilde{Q}(\mathbf{w}^{\mu_\vep})(\re^{c_1(\mathbf{w}^{\mu_\vep})}, \dots, \re^{c_n(\mathbf{w}^{\mu_\vep})})^\tau
 \ge |\Omega| P^{-1}R\mathbf{1}
-\frac{4PR}{\lambda}{\rm diag}\left\{\alpha_{11},\dots, \alpha_{nn}\right\}\mathbf{b}.\label{a73}
\eer

Noting that all the entries of $Q^{-1}$  are positive,  using   \eqref{a73}   and \eqref{a69}, we have
\ber
 &&(\re^{c_1(\mathbf{w}^{\mu_\vep})}, \dots, \re^{c_n(\mathbf{w}^{\mu_\vep})})^\tau\nn\\
  &&\ge |\Omega| \tilde{Q}^{-1}(\mathbf{w}^{\mu_\vep})P^{-1}R\mathbf{1}
   -\frac{4\tilde{Q}^{-1}(\mathbf{w}^{\mu_\vep})PR}{\lambda}{\rm diag}\left\{\alpha_{11},\dots, \alpha_{nn}\right\}\mathbf{b}\nn\\
 &&\ge (1-\vep){Q}^{-1}P^{-1}R\mathbf{1}-\frac{8{Q}^{-1}PR}{\lambda |\Omega|}{\rm diag}\left\{\alpha_{11}, \dots, \alpha_{nn}\right\}\mathbf{b}\nn\\
 &&=(1-\vep)\mathbf{1}-\frac{8{Q}^{-1}PR}{\lambda |\Omega|}{\rm diag}\left\{\alpha_{11},\dots,  \alpha_{nn}\right\}\mathbf{b}.\label{a74}
\eer

Hence from \eqref{a74} we get
 \be
\ito \left(1-\re^{c_i(\mathbf{w}^{\mu_\vep})}\re^{u_i^0+w_i^{\mu_\vep}}\right)\ud x
\le |\Omega|\vep+\frac{8}{\lambda}\sum\limits_{j=1}^n(Q^{-1})_{ij}P_jR_jb_j\alpha_{jj}, \quad  i=1, \dots, n.\label{a75}
\ee

Using  \eqref{a37c} and \eqref{a75},  we see that there exists a constant $C_\vep$ depending only on $\vep$ such that
\ber
J(\mathbf{w}^{\mu_\vep})\le \frac{|\Omega|\lambda\vep}{2}\sum\limits_{j=1}^n\frac{R_i}{P_i}+C_\vep. \label{a76}
\eer

Then it follows from Lemma \ref{lem4} and \eqref{a76} that
 \ber
 J(\mathbf{w}^{\mu_\vep})-   \inf\limits_{\mathbf{w}\in\partial\mathcal{A}} J(\mathbf{w})\le \frac{|\Omega|\lambda}{2}\left(\sum\limits_{j=1}^n\frac{R_i}{P_i}\vep-\min\limits_{1\le i\le n}\left\{\frac{R_i}{P_i}\right\}\right)
 +C\left(\sqrt{\lambda}+\ln\lambda+1\right), \label{a77}
\eer
where $C>0$ is a constant independent of $\lambda$. Now  taking $\vep$ suitably small and $\lambda$ sufficiently large
in \eqref{a77}, we get the lemma.

Therefore by Lemma \ref{lem3} and Lemma \ref{lem5} we conclude that there exists  a  large $\lambda_1>\max\{\lambda_0, \lambda_0'\}$ such that, for all
$\lambda>\lambda_1$, the functional $J$  has a minimizer in the interior of $\mathcal{A}$, say
 \ber
 \mathbf{w}^\lambda\in {\rm int}\mathcal{A}. \label{a78}
\eer

It is straightforward  to check that
  \ber
\mathbf{v}^\lambda=\mathbf{w}^\lambda+\mathbf{c}(\mathbf{w}^\lambda)\label{a79}
\eer
is a critical point of $I$  and accordingly  a solution of the system \eqref{a15}. Hence the second conclusion of Theorem \ref{th1} follows.

\section{Asymptotic behavior and quantized integrals}
\setcounter{equation}{0}
\setcounter{lemma}{0}
 In this section we prove the last two conclusions of Theorem \ref{th1}. We first establish the asymptotic behavior of the solution obtained above as $\lambda\to \infty$.
\begin{lemma}\label{lem6}
  Let $\mathbf{v}^\lambda$  be given by  \eqref{a79}. Then,
  \ber
 \lim\limits_{\lambda\to\infty}\ito (\re^{u_i^0+v_i^\lambda}-1)^2\ud x=0, \quad i=1, \dots, n.\label{a80}
\eer
\end{lemma}

{\bf Proof.} For any $\vep\in (0, 1)$,   we conclude from \eqref{a76} that there exist constants $\lambda_\vep>0$ and $C_\vep>0$
such that
 \ber
 J(\mathbf{w}^\lambda)=\inf\limits_{\mathbf{w}\in\mathcal{A}}J(\mathbf{w})\le \frac{|\Omega|\lambda\vep}{2}\sum\limits_{i=1}^n\frac{R_i}{P_i}+C_\vep,\label{a81}
\eer
for all $\lambda>\lambda_\vep$.

Since $Q$ is positive definite and  $\beta_0$ being the smallest eigenvalue of $Q$, we get
 \ber
  \ito (\mathbf{U}-\mathbf{1})^\tau Q(\mathbf{U}-\mathbf{1})\ud x\ge \beta_0\sum\limits_{i=1}^n\ito(\re^{u_i^0+v_i}-1)^2\ud x.\label{a82}
\eer
Estimating $c_i$ as that in Lemma \ref{lem3} and using \eqref{a82}, we see that
 \ber
 J(\mathbf{w}^\lambda)\ge\frac{\beta_0\lambda}{2}\sum\limits_{i=1}^n\ito(\re^{u_i^0+v_i^\lambda}-1)^2\ud x-C(\ln\lambda+1)\label{a83}
\eer
where $C>0$ is a constant independent of $\lambda$.

Therefore, we infer from \eqref{a81} and \eqref{a83} that
 \ber
 \limsup\limits_{\lambda\to\infty}\sum\limits_{i=1}^n\ito (\re^{u_i^0+v_i^\lambda}-1)^2\ud x\le \frac{\vep|\Omega|}{\beta_0}\sum\limits_{i=1}^n\frac{R_i}{P_i}, \quad \forall\, \vep\in(0, 1).
\eer
Since $\vep\in(0, 1)$ is arbitrary, the lemma follows immediately.

Then using the translation \eqref{a7} and   Lemma \ref{lem6}   we get the third conclusion of Theorem \ref{th1}.

Finally, we can establish  the quantized integrals \eqref{a5c}.  In fact, for the obtained solution, integrating the equations \eqref{a14},  we see that
desired quantized integrals follow.

The proof of Theorem \ref{th1} is now complete.

\medskip
\medskip

\small{Han  was supported in part by the Natural Science Foundation of China under grants 11201118  and by the Key foundation for Henan colleges under grant 15A110013. Both Han and Yang were supported in part by the Natural Science Foundation of China under grants 11471100.   }

\small{

}
\end{document}